\documentclass[letterpaper,final,authoryear,3p,twocolumn]{elsarticle} %11pt, %,times
\pdfoutput=1
\usepackage{amssymb,amsthm}
\usepackage{balance}

\bibpunct{(}{)}{,}{a}{}{,}
\usepackage{amsmath,bm,url}

\newcommand{\pder}[2]{\frac{\partial #1}{\partial #2}}
\newcommand{\dr}[2]{{\mathrm d}#1/{\mathrm d}#2}
\newcommand{\pdr}[2]{\partial #1/\partial #2}
\newcommand{\dd}{\,{\mathrm d}}

\newcommand{\Var}{\mathop{\mathgroup\symoperators Var}\nolimits}

  \makeatletter
  \def\ps@pprintTitle{%
     \let\@oddhead\@empty
     \let\@evenhead\@empty
     \def\@oddfoot{\footnotesize\itshape
       \@journal}%
     \let\@evenfoot\@oddfoot}
  \makeatother

\usepackage[pdfborder={0 0 0},breaklinks]{hyperref}
\journal{Manuscript in review. \hfill June 2021}

\begin{document}

\begin{frontmatter}

\title{Plasticity as a link between spatially explicit, distance-independent,\\
 and whole-stand forest growth models}

\author{Oscar Garc\'{\i}a}

\address{\href{https://orcid.org/0000-0002-8995-1341}{ORCID: 0000-0002-8995-1341}\\
    \url{garcia@dasometrics.net}} % Dasometrics, Conc\'on, Chile}
%\ead{garcia@dasometrics.net}
%\fntext[]{ORCID: 0000-0002-8995-1341}

\begin{abstract}
Models at various levels of resolution are commonly used, both for forest management and in ecological research. They all have comparative advantages and disadvantages, making desirable a better understanding of the relationships between the various approaches. It is found that accounting for crown and root plasticity creates more realistic links between spatial and non-spatial models than simply ignoring spatial structure. The article reviews also the connection between distance-independent models and size distributions, and how distributions evolve over time and relate to whole-stand descriptions. In addition, some ways in which stand-level knowledge feeds back into detailed individual-tree formulations are demonstrated. The presentation intends to be accessible to non-specialists.

\textbf{Study implications:}
Introducing plasticity improves the representation of physio-ecological processes in spatial modelling. Plasticity explains in part the practical success of distance-independent models. The nature of size distributions and their relationship to individual-tree and whole-stand models are discussed. I point out limitations of various approaches and questions for future research.
\end{abstract}


\end{frontmatter}

%\setcounter{tocdepth}{4}
%\tableofcontents

\section*{Introduction}
\label{sec:intro}

According to the level of detail, forest growth models are classified into three types \citep{weiskittel11,burkhart12,makela-valentine}: (1) individual-tree distance-dependent, or spatial, where the state of a stand is specified by sizes and spatial coordinates of every tree in a sample, (2) individual-tree distance-independent, or non-spatial, which do not use coordinates, and (3) whole-stand models, where the state is described by aggregate stand-level variables such as top height, trees per hectare and basal area. It is easier to invoke physio-ecological mechanisms in the formulation of the more detailed models. On the other hand, aggregated models, although more abstract, have the potential for producing more accurate forecasts. Therefore, there has been interest in linking the different description levels to exploit their relative strengths. This implies both mathematically deriving less detailed models from the more detailed ones, and using stand-level knowledge to guide individual-tree formulations. Attempted syntheses have generally used mean-field approximations that ignore spatial structure and individual variability \citep[e.g.][]{daniels88,picard04}. Here I explore the role of crown and root plasticity, and the use of a perfect plasticity approximation as a more realistic alternative for linking description levels.

The next section reviews relevant aspects of spatial models. It focuses on models motivated by ecological mechanisms, ignoring individual-tree models based on empirical competition indices. Conventional spatial models use lower-stem coordinates, considering the tree as a rigid radially symmetric unit. In reality, competing crowns and roots are displaced by leaning and differential growth in the direction of less contested spaces (plasticity). The implications are discussed in the \emph{Plasticity} section. When displacement is not constrained (no large gaps), in the perfect plasticity limit, the stem coordinates become irrelevant and the model becomes distance-independent. The section that follows, \emph{Stand-level Implications}, relates a perfect plasticity model to stand-level observations on canopy depth and gross volume increment. The \emph{Distributions} section connects distance-independent models to size distributions, which in turn determine whole-stand variables. It includes a brief review of how tree growth equations determine the evolution of distributions over time. It is important to realize that traditional size distribution applications only make sense as large-area limits, where tree interactions and spatial correlations can be neglected. The article ends with \emph{Discussion and Conclusions}, summarizing the main points and highlighting several open research questions.

The presentation intends to be compact and accessible to non-specialists. Mathematical derivations can be skipped without missing much. To simplify, only intraspecific competition is considered, ignoring complications arising from species mixtures. More details can be found in ``A Gentle Guide to Fully Spatial Models'', part of the documentation for the \emph{siplab} simulation package at \url{https://CRAN.R-project.org/package=siplab}. To avoid repetitive and distracting citations, refer to that report for pointers to the literature unless indicated otherwise.

\section*{\centering Spatial Models}
\label{sec:ibms}

Trees compete above ground and/or below ground for one or more limiting resources: light, water, nutrients. Resources are commonly assumed to be available at a uniform rate per unit area, so that competition can be expressed in terms of horizontal area or \emph{growing space} \citep{oliver96,ashton18}. % p.~384 [][Ch.~2,Ch.~17]
Most mechanistic individual-tree growth models can be described through an \emph{influence function} that represents the relative competitive strength of a tree of a certain species and size as a function of distance from the tree location. The resource available in any small area is allocated to the competing trees depending on their influence values at that point. In turn, the total resource captured by each tree determines its growth rate and mortality risk.

\begin{figure}[htbp]
    \centering\includegraphics[width=0.7\columnwidth]{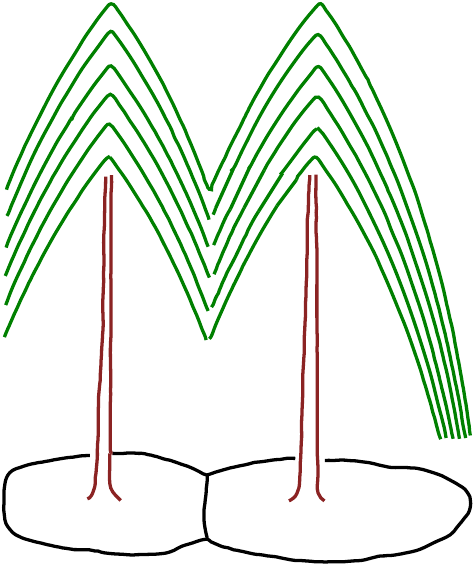}
    \caption{\sf TASS model. Foliage layers accumulate causing the crown to move up, maintaining its shape (after \citet{mitchell75}).}\label{fig:fig1}
\end{figure}

An example that is easy to visualize is the TASS model of \citet{mitchell75}. This model has been highly influential, and versions of it are still in use \citep{weiskittel11, makela-valentine,strigul08}.
TASS assumes that branch growth in length is
proportional to height increment, decreasing with distance from the top.
Every year, a new layer of foliage forms near the tip of the branches.
Foliage eventually dies at a certain depth into the canopy, where losses from respiration exceed gains from photosynthesis. Consequently, the
live crown moves up with height growth, maintaining a constant shape (Fig.~\ref{fig:fig1}).
Branch growth stops on contact with neighboring trees.
Tree stem volume growth depends on the amount of intercepted light, proportional to a depth-weighted amount of live foliage computed by
numerical integration on a 3-dimensional grid.

\begin{figure}[htbp]
    \centering\includegraphics[width=\columnwidth]{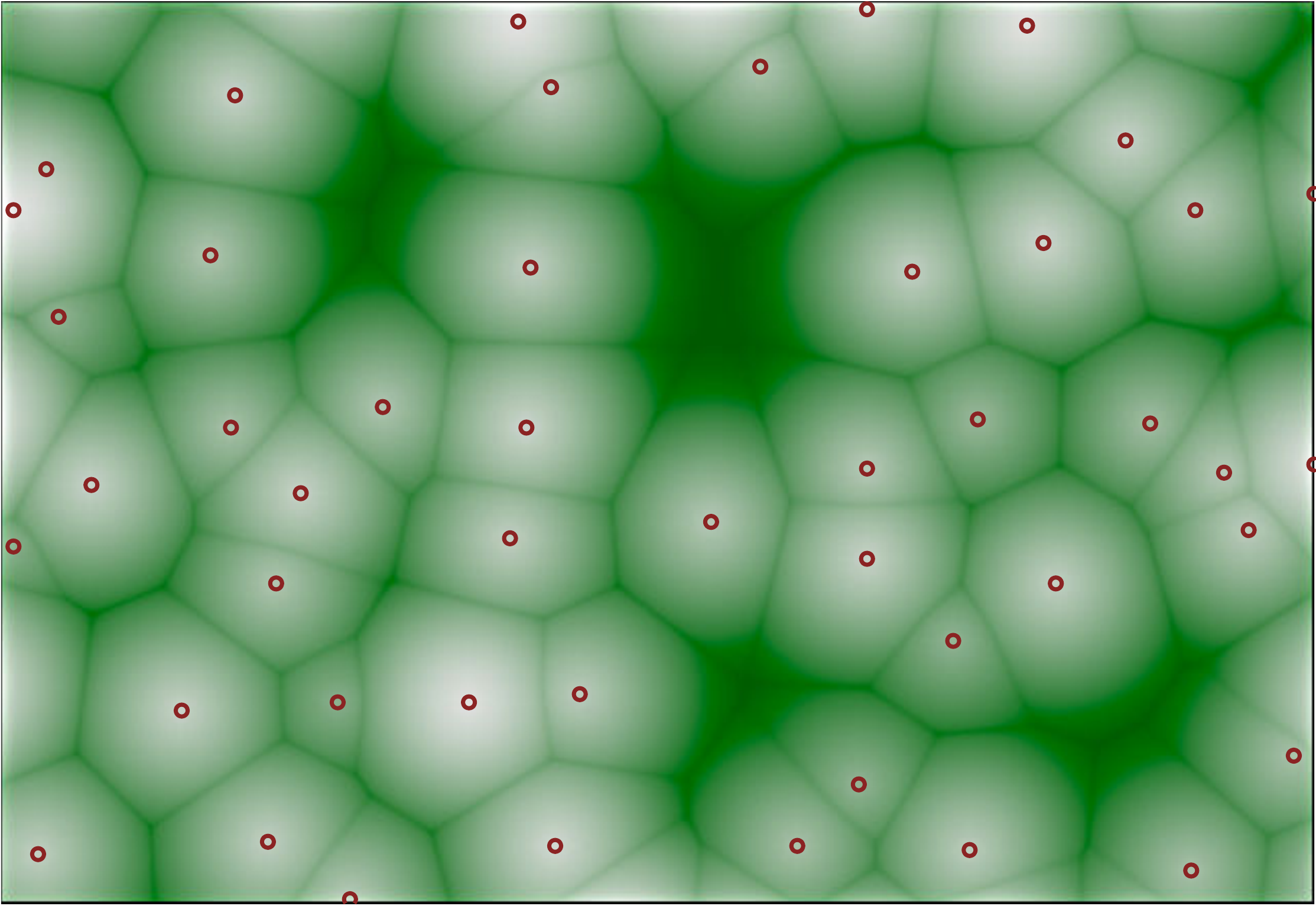}
    \caption{\sf TASS model. Top view, growing-space tessellation. Based on the \emph{spruces} data set from the \emph{spatstat R} package. Shows only a 32 $\times$ 22 m central portion of the full 56 $\times$ 38 m plot, to avoid edge effects. The tessellation is an example of generalized additively weighted Voronoi diagram \citep[][p.~126]{okabe}.}\label{fig:fig2}
\end{figure}

A simpler approach gives an equivalent result for closed canopies. Consider as an influence function the height of the upper surface of the potential crown (the crown for a tree without competitors). The actual crown extends over the area where the tree influence is higher than that of the other trees, and the light interception is proportional to that horizontal projection area (Figs.~\ref{fig:fig1}, \ref{fig:fig2}). On free crown edges, this would overestimate the amount of foliage and light interception, although the effect is likely to be small except for very young trees --- see the rightmost edge in Fig.~\ref{fig:fig1}.

The model can be generalized in several ways:
\begin{enumerate}
\item In many forest types, the crowns do not interlock or approach each other. For instance, boreal forests typically exhibit \emph{crown shyness}, with wide gaps between neighboring crowns. In those instances, influence functions can be interpreted more abstractly as a competitive strength or shading potential that extends beyond the physical crown limits. The influence function may also represent below-ground competition for water or nutrients, or a combination of above- and below-ground competition. 
\item In TASS, competition is \emph{one-sided} or \emph{completely asymmetric}: at any point on the plane, the tree with the highest influence captures all the resource available at that spot. On the contrary, one could assume that the resource is shared among the trees, for instance, in proportion to their local influence function values. Then there is no well-defined tessellation of the growing space, making visualization somewhat less intuitive, and some mathematical derivations become more difficult. Below-ground competition is likely to be more symmetric than competition for light. Also, one-sided light allotment does not allow trees to survive under the canopy, % (unless they have lower foliage light requirements than canopy trees),
which may not be realistic in shade-tolerant species. Note that this concept of local symmetry/asymmetry differs from the usual one \citep{weiner90}, in that it applies to points and not to whole plants.
\item The benefit of a unit of resource may decrease with distance. Reaching distant areas can imply a higher expense in energy and materials (branches, roots). Then, when integrating the captured resources it may be reasonable to weigh by a distance-dependent \emph{efficiency function}, what ecophysiologists call \emph{resource use efficiency} (RUE).
\item Often, the availability of moisture and nutrients is not uniform. Such heterogeneity can induce positive spatial correlations in tree sizes. The effects could be simulated through a resource availability map, although I am not aware of any published examples.
\end{enumerate}
All these extensions are implemented in the \emph{siplab} package. Apart from TASS, other models can be cast in this framework, including those based on EFT and FON approaches \citep[][Sect.~9.2.2.6]{burkhart12}. These have used a variety of influence shapes and local symmetry assumptions.

\section*{\centering Plasticity}
\label{sec:plast}

Spatial individual-tree models normally assume a rigid radially-symmetric influence function, centered on stem-base or breast-height coordinates. In reality, phototropism causes differential branch growth, foliage redistribution, and stem leaning, displacing competing crowns to occupy less contested areas --- crown plasticity. Something similar occurs with roots below ground. Consequently, locations are less important, and full canopy closure occurs earlier and at lower stand densities than predicted by models that ignore plasticity.

\begin{figure*}[htbp]
    \centering\includegraphics[width=\textwidth]{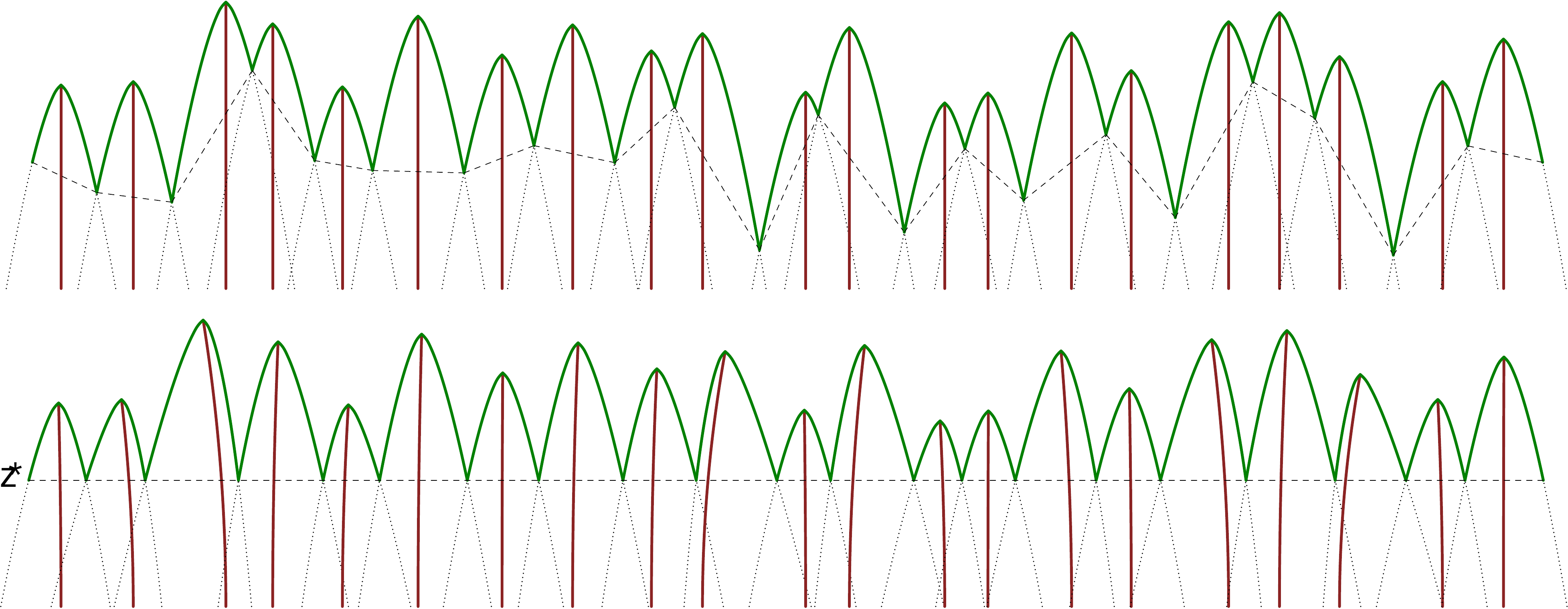}
    \caption{\sf Top: Crown profiles or influence functions in a spatial model with no
plasticity. Bottom: Leaning and shape distortion equalizing neighbors' competition intensity.}\label{fig:fig3}
 \end{figure*}

A plausible assumption is that trees tend to balance competitive pressure on opposite sides, resulting in a more uniform height $z^*$ for the influence function intersection points (Fig.~\ref{fig:fig3}). In three dimensions, with radially symmetric influence functions (circular horizontal cross-sections), it is not possible to achieve an exactly uniform $z^*$, but plasticity tends to minimize its variability \citep{strigul08}. Alternatively, one could allow the cross-sections to deform into a polygonal shape at the intersection level $z^*$, specifying the influence function more loosely through its cross-sectional area as a function of distance from the top: $A = f(h - z)$.

If there are no limits to the displacements, one has the \emph{perfect plasticity approximation}, PPA \citep{strigul08}. With full canopy closure and one-sided competition, ignoring variation of influence intersection levels around a mean $z^*$, the growing space for tree $i$ is
\begin{equation} \label{eq:A}
    A_i = f_i(h_i - z^*) \;.
\end{equation}
The sum over all the trees must equal the total area, giving the mean growing space
\begin{equation} \label{eq:z}
    \overline A = \overline{f_i(h_i - z^*)} = 1/N \;,
\end{equation}
where $N$ is the number of trees per unit area. This equation can be solved for $z^*$, in general numerically. The growing space of any tree can then be calculated from its size. One can also obtain efficiency-weighted assimilation indices. For symmetric competition, analogous explicit results are not currently available, and growing space and assimilation can only be approximated by simulation.

\begin{figure}[htbp]
    \centering\includegraphics[width=\columnwidth]{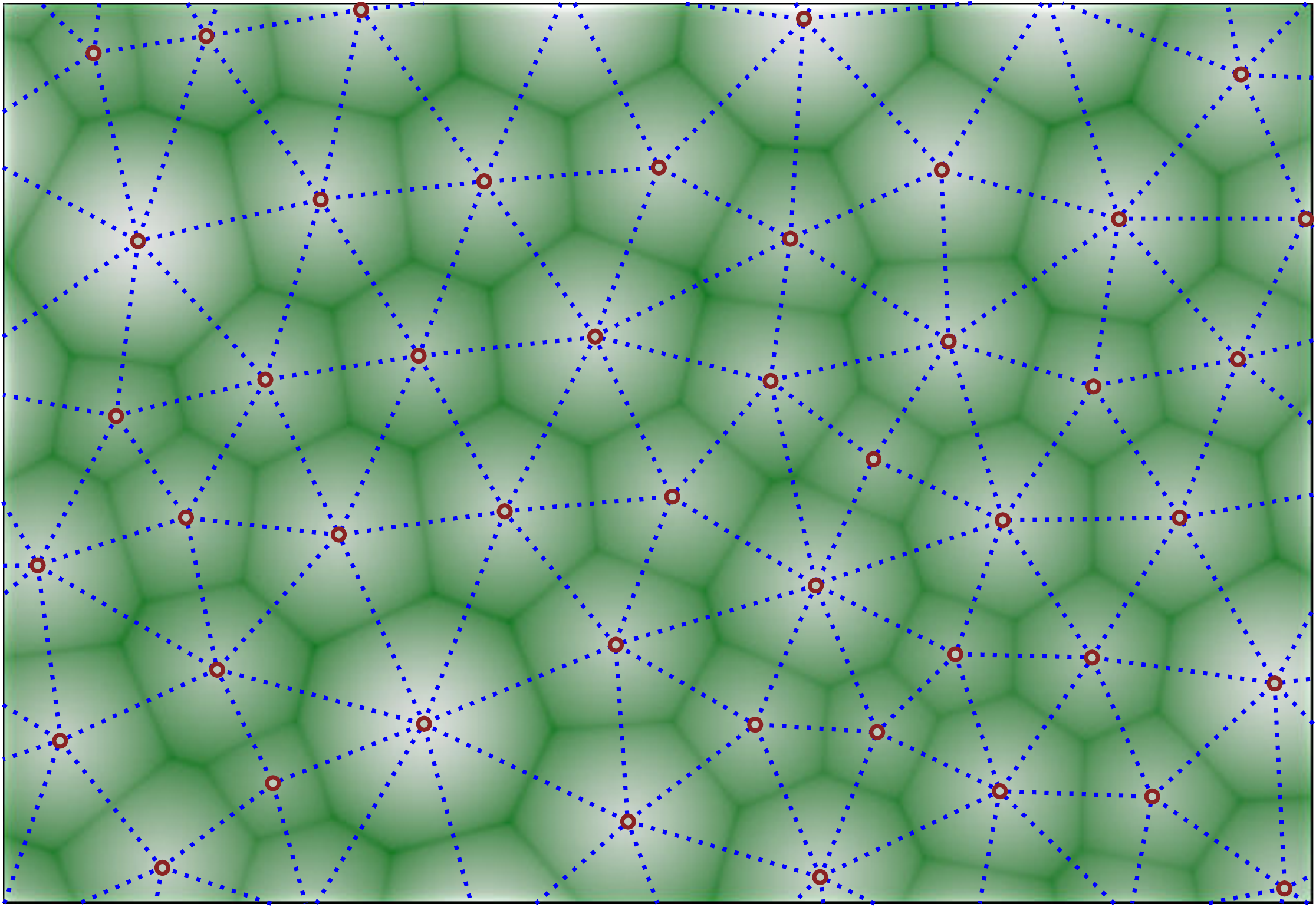}
    \caption{\sf The trees from Fig.~\ref{fig:fig2} with simulated plasticity. Dashed lines join direct competitors, forming a planar triangulation.}\label{fig:fig4}
\end{figure}

In \emph{siplab}, plasticity can be simulated by iteratively moving the location of each influence function to the centroid of the current available area (Fig.~\ref{fig:fig4}). The procedure seems to converge to the proper solution, although the correctness of the algorithm has not been mathematically proven. \citet{du06} give a proof for a problem equivalent to the case where all trees have the same size. 
It is possible to include limits or penalties to the displacements from the original tree location. 

There are limits and costs to crown or root displacements, and the PPA would not be realistic for stands with large gaps. 
With more regular spatial patterns, however, small displacements are sufficient to fulfill the PPA assumptions (Fig.~\ref{fig:fig3}). Even for highly heterogeneous natural stands, it has been argued that assuming perfect plasticity is better than assuming no plasticity at all \citep{strigul08}.

\begin{figure}[htbp]
    \centering\includegraphics[width=\columnwidth]{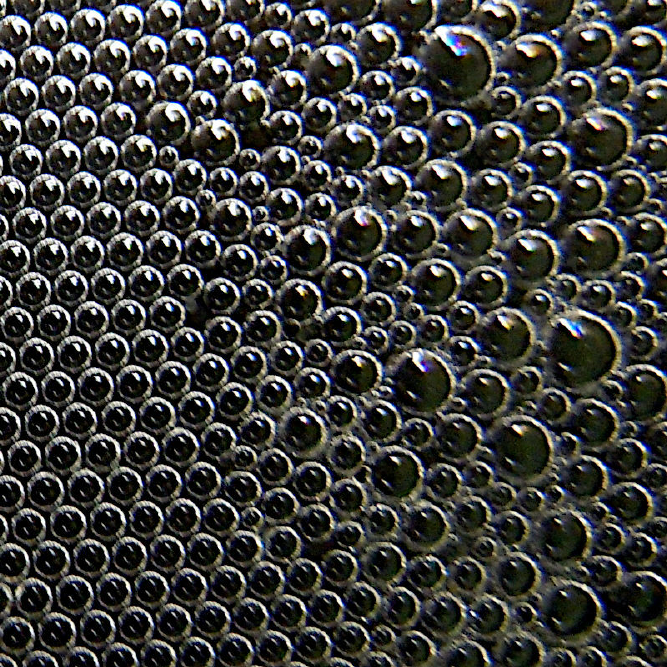}
    \caption{\sf Froth bubbles. From \protect\url{https://commons.wikimedia.org/w/index.php?title=File:Order_and_Chaos.tif&oldid=461924668}.}\label{fig:fig5}
\end{figure}

\begin{figure}[htbp]
    \centering\includegraphics[width=\columnwidth]{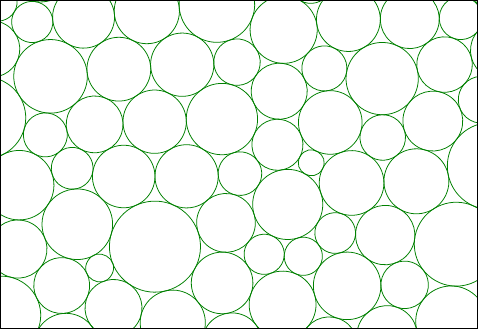}
    \caption{\sf Circle packing. Derived from the triangulation of Fig.~\ref{fig:fig4} with the Collins-Stephenson algorithm in the \emph{R} package \emph{packcircles}.}\label{fig:fig6}
\end{figure}

Beyond specific details, the importance of the perfect plasticity concept lies in that tree locations become irrelevant. A useful analogy is a froth, where bubbles move in equilibrium with their neighbors (Fig.~\ref{fig:fig5}). There are also similarities with circle packing problems (Fig.~\ref{fig:fig6}). Some of the extensive mathematical work on circle packing might be applicable to competition models \citep{wikicp,stephenson05}. 

We have implicitly assumed a flat terrain. Topography can be included by using tree-top altitude instead of height above ground \citep{umeki95b}. The result is a down-slope displacement of the influence functions, consistent with observed tree leaning, but with little or no effect on the relative neighborhood configuration.

\section*{\centering Stand-level Implications}
\label{sec:stand}

Perfect plasticity connects some aggregated stand-level properties to tree-level characteristics. Here we examine the effects of stand density on canopy depth and biomass increment for monospecific stands with one-sided competition and a closed canopy.

Given some reasonable premises about growing space partitioning \citep{gates79}, it has been found that influence functions should be of the form
\begin{equation} \label{eq:if}
    z = h - b r^a \;,
\end{equation}
where $h$ is tree size, $r$ is radial distance from the tree location (or influence function center), and $a$ and $b$ are parameters. As tree size we use height in preference to the more commonly used stem diameter or volume, since stem thickness is unlikely to have a direct causal effect on volume or biomass growth rates. Equation \eqref{eq:if} is fairly flexible. For instance, $z = h - 2 r^{1.3}$ is a close approximation to the TASS potential crown profile $z = h - 6.1 [\exp(r / 3.432) - 1]$.

From eq.~\eqref{eq:if}, the cross-sectional area $\pi r^2$ of the influence surface at a level $z$ is $\pi [(h-z)/b]^{2/a} \equiv c (h-z)^{2/a}$. Therefore, as in the previous section, with perfect plasticity the growing space of tree $i$ is
\begin{equation} \label{eq:gs}
    A_i = c (h_i - z^*)^{2/a} \;,
\end{equation}
where $z^*$ satisfies
\begin{equation} \label{eq:z*}
    c \overline{(h_i - z^*)^{2/a}} = 1/N \;.
\end{equation}
In general, there is no closed-form solution, but the mean can be approximated using a Taylor expansion
\begin{align*}
    x^k &\approx \overline{x}^k + k \overline x^{k-1}(x - \overline x) 
        + \frac{k (k-1)}{2} \overline x^{k-2}(x - \overline x)^2 \\
    \overline{x^k} &\approx \overline{x}^k + \frac{k (k-1)}{2} \overline x^{k-2} \Var[x] \;,
\end{align*}
or
\begin{equation} \label{eq:mm}
    \overline{x^k} \approx \left(1 + \frac{k (k-1)}{2} C^2\right) \overline{x}^k \;, 
\end{equation}
where $C$ is the coefficient of variation.
% This result is exact for $k=1$, for $k=2$, and for any $k$ if $x$ has a Normal distribution (because then the neglected higher moments are 0).
Applying this to eq.~\eqref{eq:z*} it is found that, if $C$ is relatively stable, $\overline h - z^*$ is approximately proportional to $1 / N^{a/2}$.

The mean canopy depth is $\overline h - z^* + d$, where $d$ is the thickness of the foliage layer, see Figs.~\ref{fig:fig1} and \ref{fig:fig3}. Therefore, it approximates a linear function of $1 / N^{a/2}$. Studies in forest plantations have produced straight lines for canopy depth over the average spacing $1/\sqrt N$ \citep[][sect.~2.4.1]{makela-valentine}. That suggests conical influence functions ($a = 1$) or, more generally, functions where the area increases as the square of distance from the top (eq.~\eqref{eq:gs}).

The annual increment in biomass or stem volume of a tree $i$ could be assumed to be proportional to its growing space $A_i$. Then, the stand gross increment per unit area (including any trees that might die, also known as \emph{gross growth}, \emph{accretion}, or \emph{net primary production}) would be proportional to the sum of the $A_i$ divided by the stand area. After canopy closure, this ratio is 1 (Figs.~\ref{fig:fig2}, \ref{fig:fig4}, and eq.~\eqref{eq:z}). Therefore, the gross increment in closed-canopy stands, for a given site quality, would be constant, regardless of stand density. This is known to be a good first approximation \citep[e.g.][p.~354]{hawley54}, %ashton18, Ch.17
but some dependency on stand density has been observed.

\begin{figure}[htbp]
    \centering\includegraphics[width=\columnwidth]{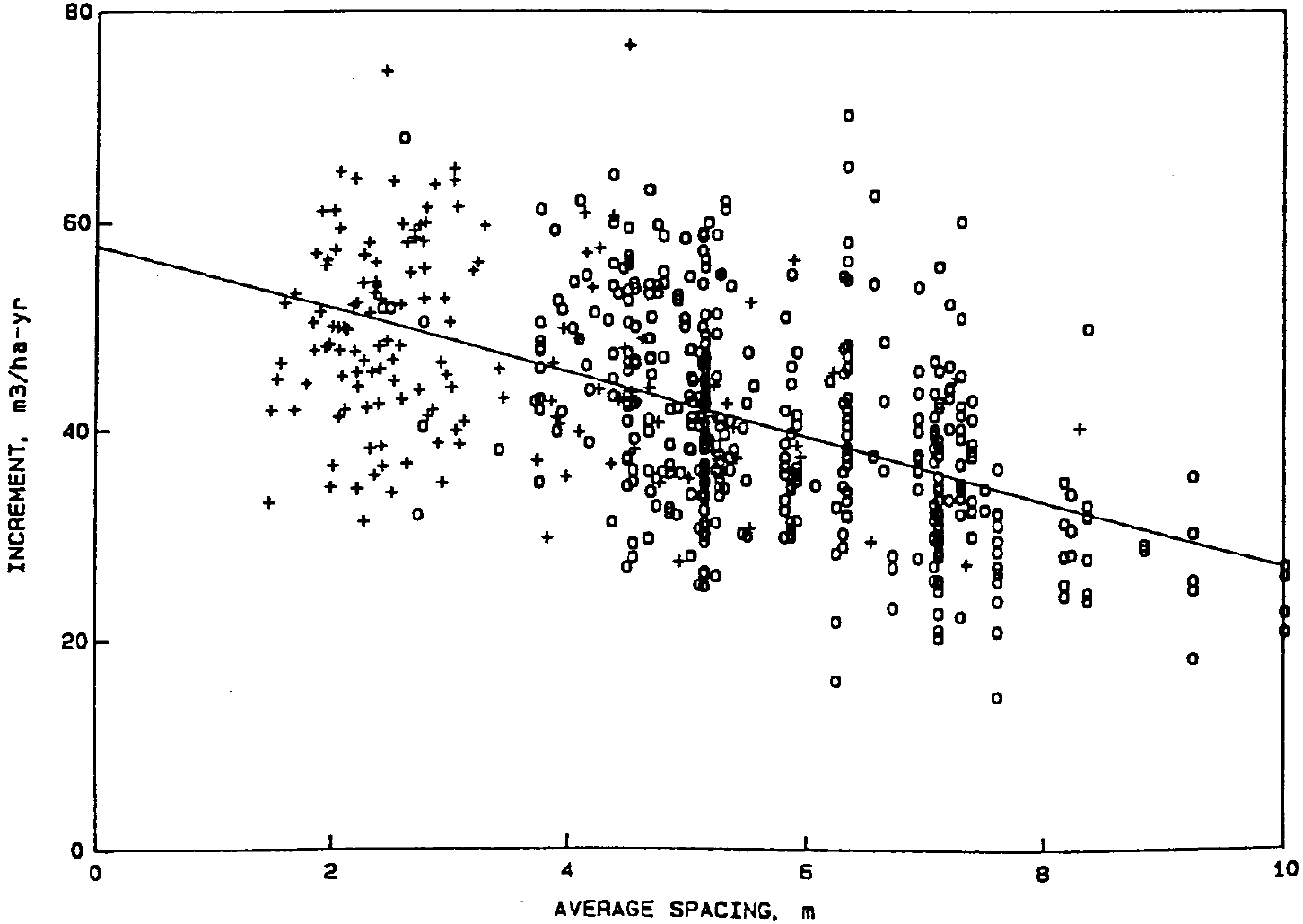}
    \caption{\sf Closed-canopy annual gross volume increment per hectare in radiata or Monterey pine plantations (\emph{Pinus radiata} D. Don), adjusted for site quality. \emph{Average spacing} is $1/\sqrt N$, for $N$ in trees per square meter. The symbol $+$ indicates measurements that include mortality. From \citet{thinned}.}\label{fig:fig7}
\end{figure}

Figure \ref{fig:fig7} shows gross volume increments over an unusually wide range of densities.
The effect of stand density can be explained if resource use efficiency decreases with distance. Total effective assimilation would then decrease with increased spacing. Specifically, consider an efficiency function $E(r) = 1 - p r^q$, where $r$ is radial distance from the center of an influence function, and $p$ and $q$ are parameters. 
To simplify, let us neglect out-of-roundness and take the growing space of tree $i$, with perfect plasticity and one-sided competition, as a circle with area $A_i = \pi R_i^2$. The efficiency-weighted area $A'_i$ can be obtained by adding over narrow rings of radius $r$ and thickness $\dd r$. In the limit $\dd r \to 0$,
\begin{align*}
    A'_i =& \int_0^{R_i} E(r) (2\pi r) \dd r = 2 \pi \int_0^{R_i} (1 - p r^q) r \dd r \\
         =& \pi R_i^2 - \frac{\pi p}{q + 2} R_i^{q + 2} \;,
\end{align*}
from where
\begin{equation} \label{eq:Aprime}
    A'_i = A_i - \frac{p}{\pi^{q/2}(q + 2)} A_i^{q/2+1} \;.
\end{equation}
The growth rate is proportional to the total weighted area
\[
    \sum A'_i = N \overline{A'} = N \overline A - \frac{p}{\pi^{q/2}(q + 2)} N \overline{A^{q/2+1}} \;.
\]
Using eq.~\eqref{eq:mm},
\begin{multline*}
    \sum A'_i \approx N \overline A - \\
     \frac{p}{\pi^{q/2}(q + 2)}\left(1 + \frac{k (k-1)}{2} C^2\right) N \overline{A}^{q/2+1} \;,
\end{multline*}
and from eqs.~\eqref{eq:gs}--\eqref{eq:z*},
\begin{multline} \label{eq:sumAprime}
    \sum A'_i \approx 1 - \\
     \frac{p}{\pi^{q/2}(q + 2)}\left(1 + \frac{k (k-1)}{2} C^2\right) N^{-q/2} \;.
\end{multline}
Therefore, provided that the coefficient of variation of $A_i$ does not change much with $N$, the gross volume or biomass increment in closed stands tends to decrease linearly with $N^{-q/2}$. A conical efficiency function ($q = 1$) would produce the straight line of Fig.~\ref{fig:fig7}.

%The same result is obtained under more general conditions through a scaling argument: 

\section*{\centering Distributions}
\label{sec:dist}

Perfect plasticity converts a distance-dependent model into a distance-independent one: a tree growth rate depends on its size, and possibly on stand-level variables like stand density and site quality, but not on neighbor sizes or locations. E.g., see eqs.~\eqref{eq:A}, \eqref{eq:Aprime}.

\begin{figure}[htbp]
    \centering\includegraphics[width=\columnwidth]{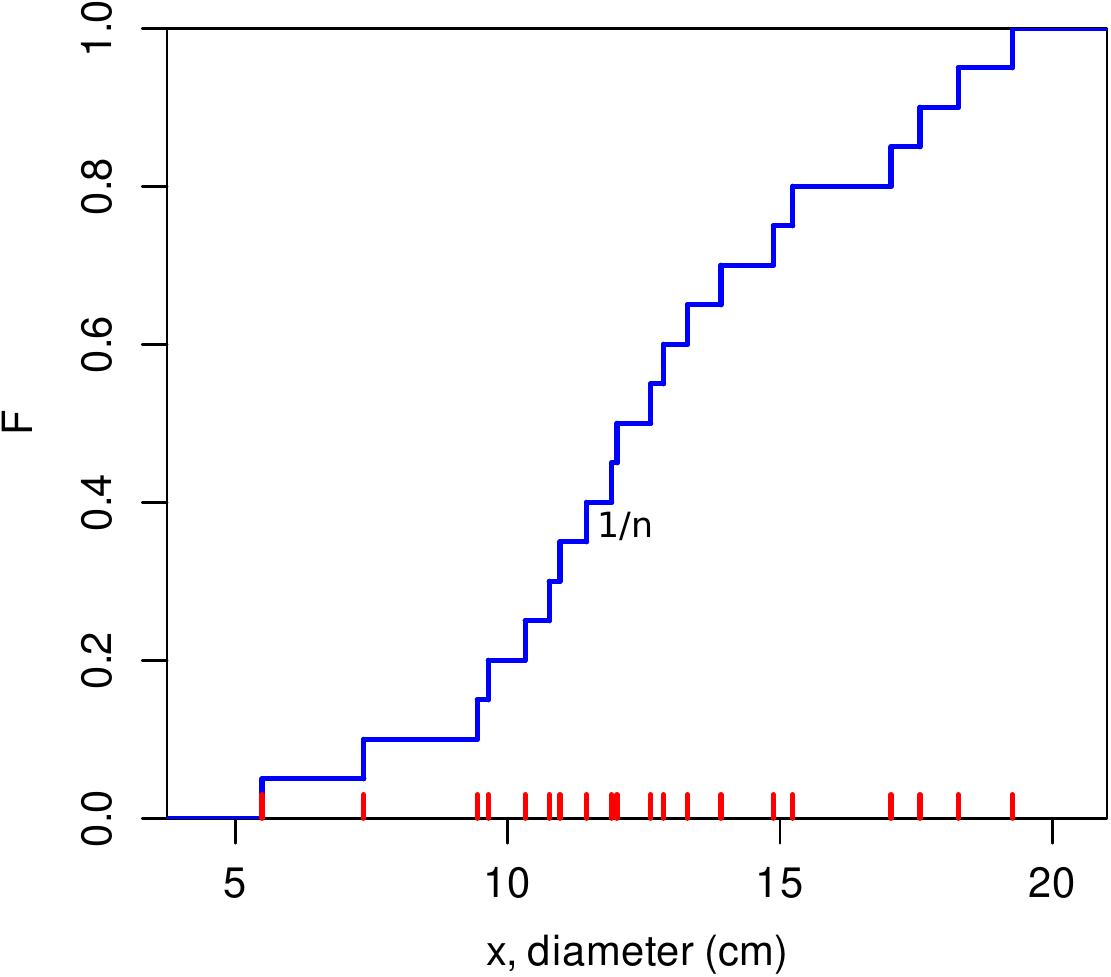}
    \caption{\sf Equivalence between tree lists and size distributions. Example with a list of $n=20$ tree diameters indicated by tics on the $x$-axis. The cumulative size distribution $F(x)$, a step function with steps $1/n$, gives the proportion of trees with diameters $\le x$ for any $x$. If the tree list had weights $w_i$, the step size would be $w_i/\sum w_j$.}\label{fig:fig8}
\end{figure}

The state in a distance-independent model is commonly specified as a \emph{tree list}, an enumeration of the sizes, usually stem diameters, of the trees in a certain area. Sometimes a weight is assigned to each size, to handle variable probability sampling and to accommodate some mortality projection methods \citep[][Sec.~5.5]{weiskittel11}. The tree list is equivalent to a (cumulative) size distribution $F(x)$ that gives the proportion of trees with size less than or equal to $x$ (fig.~\ref{fig:fig8}). As the number of trees tends to infinity, $F(x)$ becomes a continuous smooth function with a distribution density $f(x) = \dr{F(x)}{x}$. All this applies also if the ``size'' is a vector, containing for instance diameter and height, with the convention that $(d, h) \le (x, y)$ means $d \le x$ \emph{and} $h \le y$.

Tree size distributions are often treated as probability distributions, although the nature of the analogy and the meaning of the probability are rarely discussed. Considering a number of trees in a certain observation window, such as a sample plot, one might interpret the probability as a proportion of trees when the plot location varies at random over a stand or forest, a \emph{design-based} view. Alternatively, the probability may represent uncertainty about the size of a specific tree, a \emph{model-based} approach. Either way, in general it is unrealistic to take tree sizes as independently distributed: competition induces negative spatial correlations, while micro-site fertility gradients produce positive correlations \citep{fox01}. For finite observation windows, the distribution should be seen only as a marginal probability distribution. The marginal is sufficient for determining statistics that are linear on the distribution function, like means and totals. However, variances, higher moments and order statistics depend on the joint distribution, and vary with plot size \citep{fox01,sambakhe14}. As the size of the observation window tends to infinity the effect of local interactions vanishes, making the limiting continuous size distribution useful in management to predict product sizes at the stand or compartment level. It is, however, an incomplete state description for stand dynamics if tree interactions are significant.

In principle, with distance-independent growth, it is possible to derive mathematically the evolution of a size distribution, as an alternative to algorithmic tree-list methods. Let $F(x, t)$ be the size distribution in a certain area at time $t$, that is, the proportion of (live) trees with size less than or equal to $x$. For now assume that $x$ is a single variable, typically stem diameter, that growth is deterministic, and that any mortality is negligible. If $x$ grows from $x_1$ at time $t_1$ to $x_2$ at a later time $t_2$, preservation of the number of trees implies that
\begin{multline} \label{eq:FF}
    F(x_2, t_2) = F(x_1, t_1) \;, \\ %\quad
       \text{for } t_1 \le t_2 \text{ and } x_1 \le x_2 \;.
\end{multline}
Let tree growth be $\Delta x = g(x) \Delta t$, where the time increment $\Delta t$ might be finite, typically 1, 5, or 10 years in discrete-time models, or an infinitesimal in continuous-time models. It follows that
\begin{equation} \label{eq:FFdelta}
    F(x + g(x) \Delta t, t + \Delta t) = F(x, t) \;.
\end{equation}
This is valid for any number of trees.

In the large-area limit where $F$ is continuous, and in continuous time ($\Delta t \rightarrow 0$), the left-hand side of eq.~\eqref{eq:FFdelta} can be written as the first terms of a Taylor expansion, giving
\[
    F(x, t) + \pder{F(x, t)}{x} g(x) \Delta t + \pder{F(x, t)}{t} \Delta t = F(x, t) \;,
\] or \[
    g(x) f(x, t) + \pder{F(x, t)}{t} = 0 \;,
\]
where $f(x, t) = \pdr{F(x, t)}{x}$ is the distribution density. Finally, differentiating with respect to $x$, one obtains the partial differential equation (PDE)
\begin{equation*} \label{eq:liouville}
    \pder{g(x) f(x, t)}{x} + \pder{f(x, t)}{t} = 0 \;.
\end{equation*}
If $m(x)$ is the mortality rate for trees of size $x$, it is possible to account for mortality with
\begin{equation} \label{eq:vF}
    \pder{g(x) f(x, t)}{x} + \pder{f(x, t)}{t} + m(x) f(x, t) = 0
\end{equation}
\citep{picard04}. Eq.~\eqref{eq:vF} is known in physics as a continuity or Liouville equation \citep{picard01,picard04}, and in ecology as the von Foerster or McKendrick--von Foerster equation \citep{strigul08}. This PDE, together with appropriate boundary conditions (possibly including ingrowth), determines the evolution of the size distribution density over time. In general, it needs to be solved numerically, essentially going back to equations \eqref{eq:FF}, \eqref{eq:FFdelta}.

Two generalizations are important. First, if the growth $g(x)$ is considered stochastic, an additional second-order term appears in the PDE, giving the Fokker-Plank equation, also known as Kolmogorov's forward equation \citep{suzuki74}. Second, as earlier discussed, at least two variables are needed to describe adequately tree growth, e.g., height and biomass or stem volume, or height and diameter. Then, $x$ becomes a vector, and $g(x)$ should be a system of two or more equations. \citet{picard01} discuss these two extensions, although their presentation may be too technical for many readers.

More recently, stochastic differential equations (SDEs) have been found to be a more convenient alternative to Fokker-Plank equations. Instead of describing the evolution of the whole distribution, an SDE gives the changes in the mean and variance. Therefore, it would connect more directly individual-tree with whole stand models. The topic is insufficiently explored in plant ecology, but see \citet{rupsys10} and \url{https://github.com/ogarciav/resde}.

\section*{\centering Discussion and Conclusions}
\label{sec:dc}

Plasticity causes crowns, or more generally influence functions, to repel each other. The spatial distribution becomes more regular --- compare Figs.~\ref{fig:fig2} and \ref{fig:fig4}. In the perfect plasticity limit, the dependence on tree coordinates disappears, resource capture and growth depend only on tree size and stand density. Explicit relationships between resource capture and size can be obtained under one-sided competition. Equations are not yet available for less than fully asymmetric allotments, only simulated curves have been obtained.

Linear regressions of canopy depth and of gross increment \emph{vs.} average spacing have been reported in the literature. Those stand-level relationships imply conical influence and efficiency functions, at least under certain plausible modelling assumptions. Some deviations from linearity, however, cannot be ruled out with the data.

In theory, the size distributions that characterize distance-independent models can be projected over time with partial differential equations of the Liouville or von Foerster type, if tree growth is taken as deterministic. If growth is ``random'', the resultant stochastic process can be described by Fokker-Plank equations, or perhaps more conveniently by stochastic differential equations. A whole-stand model is given by distribution summaries like means and totals. The implementation details are not trivial, especially with realistic multivariate ``sizes''.

The probability analogy of size distributions assumes that the sizes in a sample are independently distributed. However, interactions among neighbors and micro-site spatial correlations generate local dependencies. Therefore, size distributions must be viewed only as marginal distributions or as large-area limits. For many purposes that is sufficient, but still, local spatial structure is clearly important for stand development and for the statistics of plot measurements. The effects of micro-site correlations, in particular, have been largely ignored in growth modelling. More research on these topics is needed.

A difficulty with size-dependent growth equations of the form $\Delta x = g(x)$, or its continuous-time equivalent, is the confounding of direct and reverse causality: larger trees may grow faster, but also faster-growing trees tend to be larger. Consequently, fit statistics can give an over-optimistic impression of the model's predictive abilities. Conventional statistical inference only measures association, and being based on an inherently symmetric probability theory, cannot discriminate causality direction \citep{pearl16,pearl18}. This is not a problem when not extrapolating outside the range of the data, but models are often used to predict under conditions that have not been previously observed.

The general framework described here, besides rising a number of unresolved research questions, can clarify connections between models at different levels of resolution. That should facilitate cross-fertilization, taking advantage of the particular strengths of the various approaches.

%\section*{Acknowledgements}
%\label{sec:ack}
%\addcontentsline{toc}{section}{Acknowledgements}

% BibTeX users please use one of
\bibliographystyle{spbasic}      % basic style, author-year citations
\balance
\bibliography{pp}   % name your BibTeX data base

%\addcontentsline{toc}{section}{References}

\end{document}